\documentclass[final,5p,times,twocolumn]{elsarticle}
\usepackage{graphics}
\usepackage{color}
\usepackage{amssymb}

\journal{Physica C}

\begin{document}

\begin{frontmatter}


\title{Angle-Resolved Photoemission from Cuprates with Static Stripes}


\author{Tonica Valla}

\address{Condensed Matter Physics and Materials Science Department, Brookhaven National Lab, Upton, NY 11973, USA}

\begin{abstract}
25 years after discovery of high-temperature superconductivity (HTSC) in La$_{2-x}$Ba$_x$CuO$_4$ (LBCO), the HTSC continues to pose some of the biggest challenges in materials science. Cuprates are fundamentally different from conventional superconductors in that the metallic conductivity and superconductivity are induced by doping carriers into an antiferromagnetically ordered correlated insulator. In such systems, the normal state is expected to be quite different from a Landau-Fermi liquid - the basis for the conventional BCS theory of superconductivity. The situation is additionally complicated by the fact that cuprates are susceptible to charge/spin ordering tendencies, especially in the low-doping regime. The role of such tendencies on the phenomenon of superconductivity is still not completely clear. Here, we present studies of the electronic structure in cuprates where the superconductivity is strongly suppressed as static spin and charge orders or \lq\lq{}stripes”\rq\rq{} develop near the doping level of $x =1/8$ and \lq\lq{}outside\rq\rq{} of the superconducting dome, for $x<0.055$. We discuss the relationship between the \lq\lq{}stripes\rq\rq{}, superconductivity, pseudogap and the observed electronic excitations in these materials. 
\end{abstract}


\end{frontmatter}


\section{Introduction}
\label{Intro}
\begin{figure}[tr]
\begin{center}
\includegraphics[width=6.5cm]{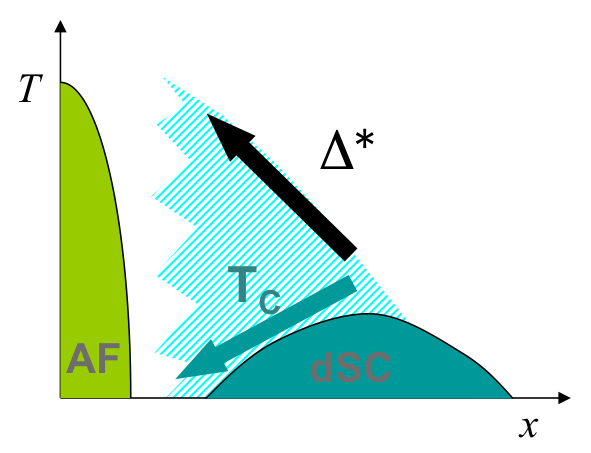}
\caption{A generic phase diagram of cuprate superconductors. In the underdoped region, the excitation gap $\Delta^{\star}$ increases while $T_c$ decreases with reduced doping}
\label{Fig1}
\end{center}
\end{figure}

\begin{figure}[]
\begin{center}
\includegraphics[width=6.5cm]{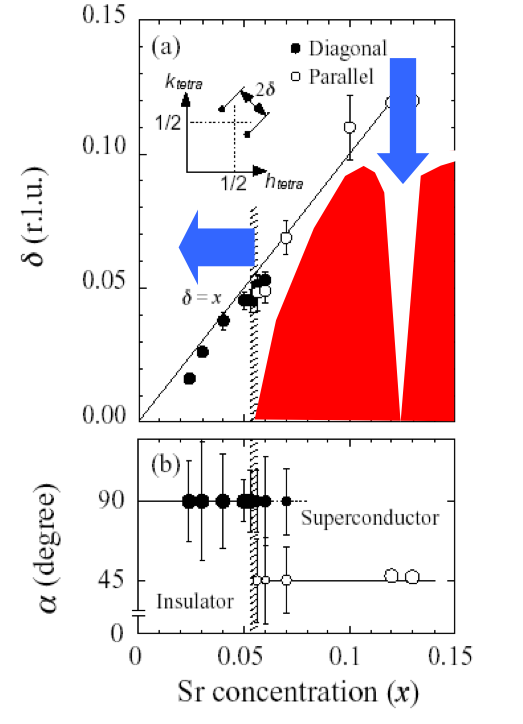}
\caption{Doping dependence of incommensurability (a), and angle (relative to the Cu-Cu direction) (b) of the  spin superstructure measured in LSCO \cite{Fujita2002}. Arrows indicate regions where the superstructure is static}
\label{Fig2}
\end{center}
\end{figure}

The underdoped side of cuprate phase diagram is full of amazing features that do not exist in conventional superconductors. One example is a normal-state gap (pseudogap), which exists above the temperature of the superconducting transition $T_c$. It is generally believed and observed that the magnitude of the pseudogap monotonically decreases with increasing doping, whereas $T_c$ moves in the opposite direction in the underdoped regime as shown in Fig. \ref{Fig1} \cite{Timusk1999,Damascelli2003}. The origin of the pseudogap and its relationship to superconductivity is one of the most important open issues in physics of HTSC and represents the focal point of current theoretical interest \cite{Franz2001,Lee2006,Emery1999,Anderson1987}. In one view, the pseudogap is a pairing (superconducting) gap, reflecting a state of Cooper pairs without global phase coherence. The superconducting transition then occurs at some lower temperature when phase coherence is established \cite{Emery1995,Uemura1989}. In an alternative view, the pseudogap represents another state of matter that competes with superconductivity. However, the order associated with such a competing state has not been unambiguously identified. Candidates were found in neutron scattering studies, where an incommensurate spin order was detected inside vortices \cite{Lake2001} and in 
scanning tunnelling microscopy (STM) experiments where a charge
ordered state, energetically very similar to the superconducting state, has been found in
the vortex cores, in the ‘pseudogap’ regime above $T_c$ and in patches of
underdoped material in which the coherent conductance peaks were absent \cite{Vershinin2004,McElroy2005,Hoffman2002}. 
These charge/spin superstructures - \lq\lq{}stripes\rq\rq{} are particularly prominent in 214 family of cuprates (LBCO and La$_{2-x}$Sr$_x$CuO$_4$ (LSCO)), where a \textit{static} spin (and charge) superstructure forms at low temperatures in the \lq\lq{}spin glass\rq\rq{} phase ($0.02<x<0.055$), and around the \lq\lq{}1/8 anomaly\rq\rq{} ($x\sim0.125$). Superconductivity is notably absent in both regimes, indicating that it competes with such orders. However, the ordering tendencies exist in most cuprates and extend well into the overdoped region. For example, neutron scattering studies show that incommensurate spin structure forms as soon as the carriers are doped into the system,  exists within the whole superconducting regime and disappears together with superconductivity at $x\sim0.3$ \cite{Wakimoto2004}. Incommensuration from the antiferromagnetic wave vector ($\pi/a,\pi/a$) grows roughly in proportion with doping till $x\sim1/8$ and tends to saturate after that. When these super-structures are static, superconductivity is either completely absent ($x<0.055$) or strongly suppressed (near $x=1/8$). In the former case the spin superstructure is diagonal, while in the latter it is parallel to the Cu-O bond \cite{Fujita2002} as shown in Fig. \ref{Fig2}. The charge order (with half-wavelength of the spin order) has been found so far only in the latter case \cite{Abbamonte2005}.  

%

\section{Parallel Stripes}
\label{Parallel}

\begin{figure}[b]
\begin{center}
\includegraphics[width=8.5cm]{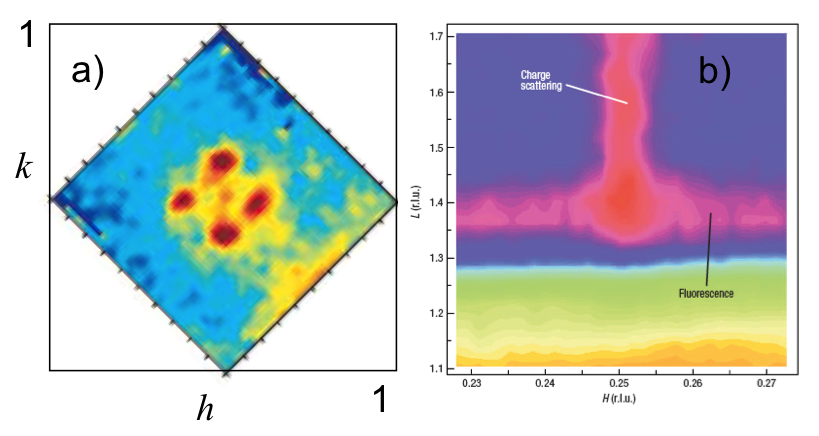}
\caption{(a) Spin \cite{Tranquada2004} and (b) charge \cite{Abbamonte2005} superstructure in LBCO at $x=1/8$.}
\label{Fig3}
\end{center}
\end{figure}

LBCO exhibits a sharp drop in superconducting transition temperature, T$_c\rightarrow0$,
when doped to $\sim1/8$ holes per Cu site ($x=1/8$), while having almost equally ‘strong’
superconducting phases, with T$_{Cmax}\sim30$ K at both higher and lower dopings \cite{Moodenbaugh1988}. Therefore, the $x=1/8$ case represents an ideal system to study the ground state of the pseudogap as the ‘normal’ state extends essentially to $T=0$. In scattering experiments (Fig. \ref{Fig3}) on single crystals, static stripes with spin period of 8 unit cells \cite{Tranquada2004,Tranquada1995} and a charge  period of 4 unit cells have been detected at low temperatures \cite{Abbamonte2005} . While superconductivity is strongly reduced at $x=1/8$, metallic behaviour seems to be preserved \cite{Adachi2001,Li2007a,Tranquada2008}. Optical studies have detected a loss of spectral weight at low frequencies with simultaneous narrowing of a Drude component, suggesting the development of an anisotropic gap \cite{Homes2006}. 

\begin{figure}[]
\begin{center}
\includegraphics[width=8.5cm]{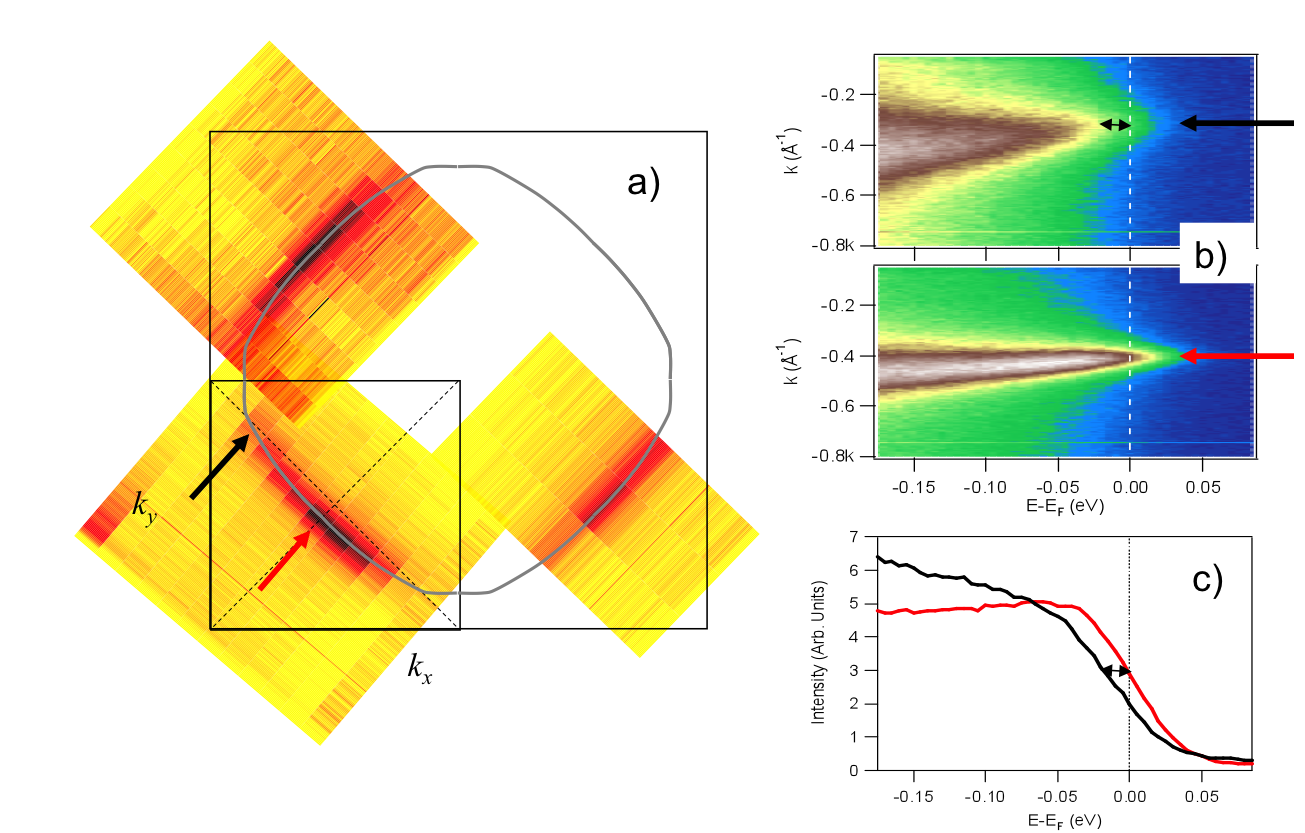}
\caption{Photoemission from LBCO at $x = 1/8$ \cite{Valla2006}. (a) Intensity at the Fermi level (Fermi surface). Arrows correspond to the momentum lines represented in (b) and (c). (b) Photoemission intensity as a function of binding energy along the momentum lines indicated in (a). (c) Energy distribution curves (EDCs) of spectral intensity integrated over a small interval around $k_F$ along the two lines in k-space shown in (b). The arrow represents the shift of the leading edge. The spectra were taken at $T =$16 K.}
\label{Fig4}
\end{center}
\end{figure}
Figure 4 shows the photoemission spectra from LBCO at $x = 1/8$ in the ordered state ($T=16$ K) \cite{Valla2006}. The momentum distribution of the photoemission intensity from the energy window of ±10 meV around the Fermi level is shown within the Brillouin zone (Fig. 4a). From these contours, the Fermi surface is extracted. The area enclosed by the Fermi line corresponds to $x = 0.115\pm 0.015$, in good agreement with the nominal doping levels, signaling that the bulk property has been probed. 
As shown in Fig. 4b and c, we have detected an excitation gap in photoemission spectra from this material with a magnitude that depends on the $k$ position on the Fermi surface, vanishing at the node and with maximum amplitude near the antinode.
\begin{figure}[]
\begin{center}
\includegraphics[width=8.5cm]{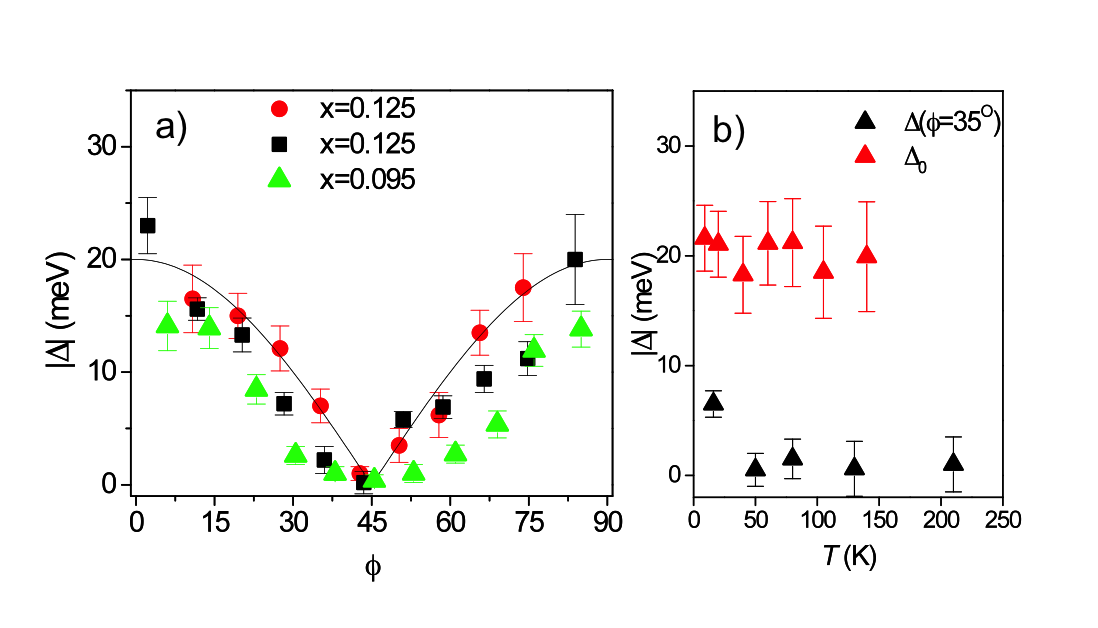}
\caption{Magnitude of single-particle gap in LBCO at $x=1/8$ and $x=0.095$ as a function of an angle around the Fermi surface at $T=16$ K (a) and as a function of temperature, for two characteristic Fermi points, for the $x=1/8$ sample (b) \cite{Valla2006}.}
\label{Fig5}
\end{center}
\end{figure}
%

 In the detailed $k$-dependence for several samples (Fig. 5a), two unexpected properties are
uncovered. First, gaps in all samples have magnitudes consistent with $d$-wave symmetry,
even though superconductivity is essentially nonexistent in LBCO at $x = 1/8$. Second, the
gap in LBCO is larger at $x = 1/8$ than at $x = 0.095$. This finding contradicts a common belief that the excitation gap in cuprates monotonically increases as the antiferromagnetic (AF) phase is approached. 

The momentum-resolved picture from ARPES is consistent with the STM data obtained from the same parent crystal used for ARPES. In Fig. 6a, a typical STM topographic image of a cleaved LBCO surface is shown. In addition, the differential conductance ($dI/dV$) spectra were taken at many points in a wide range of energies and averaged over the whole field of view. The resulting conductance spectrum (Fig.
6b) shows a symmetric V-like shape at low energies, with zero-DOS falling exactly at the Fermi level,
which is consistent with a pairing $d$-wave gap. The magnitude of this gap, $\Delta_0 \approx 20$ meV, as determined from the breaks in $dI/dV$ curve agrees with the maximal gap $\Delta_0$ measured in photoemission.
\begin{figure}[b]
\begin{center}
\includegraphics[width=6.5cm]{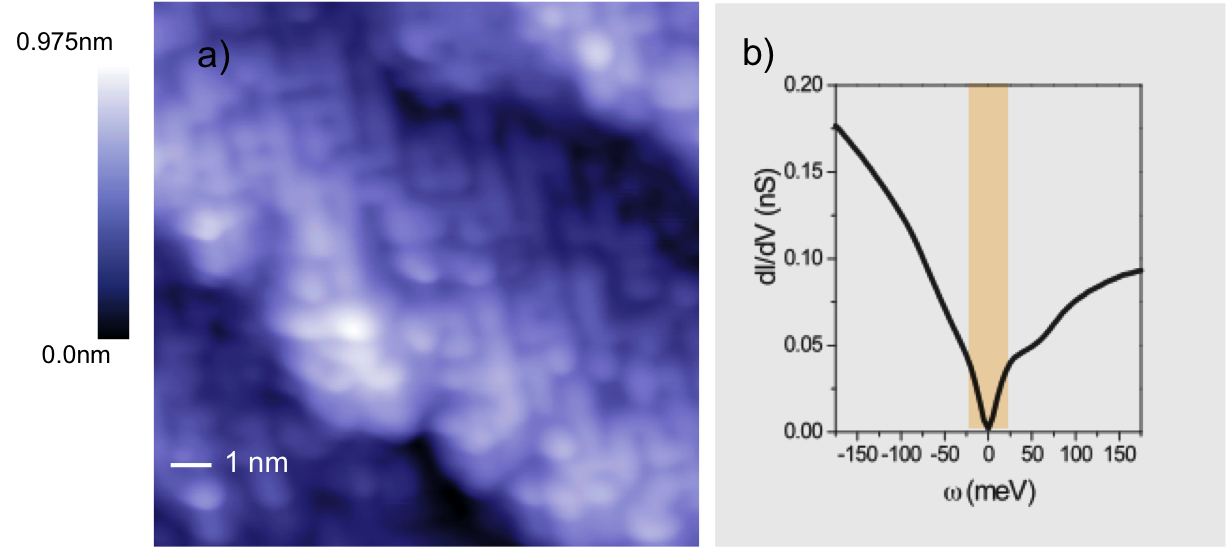}
\caption{(a) STM topographic image of the cleaved LBCO sample at 4.2 K. (b) A tunneling conductance spectrum averaged over the area shown in (a). A V-like profile of DOS for energies $|\omega|<20$ meV is consistent with a $d$-wave gap observed in ARPES \cite{Valla2006}.}
\label{Fig6}
\end{center}
\end{figure}
\begin{figure}[]
\begin{center}
\includegraphics[width=6.5cm]{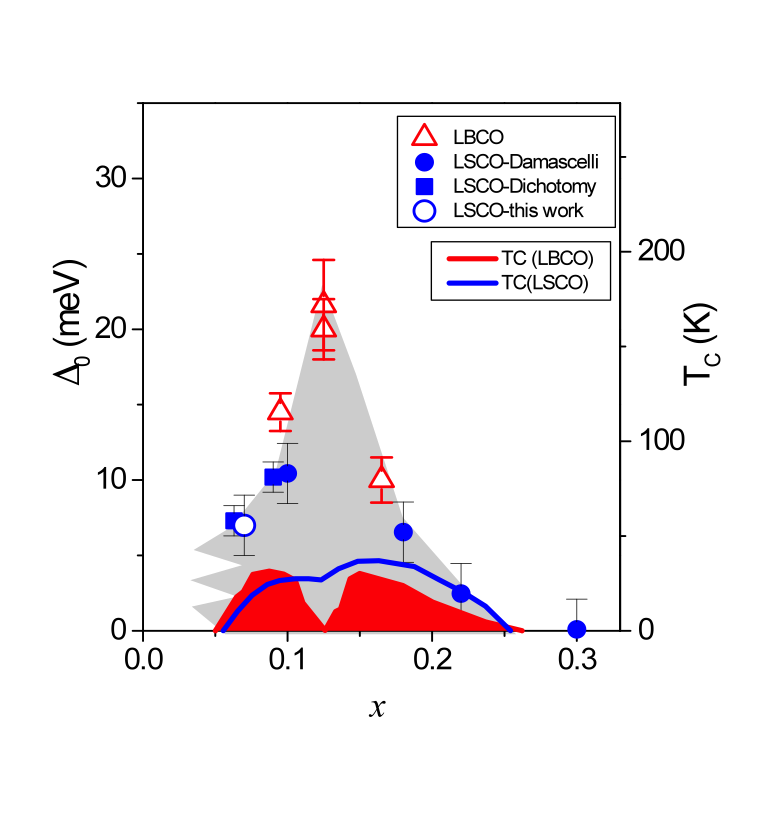}
\caption{Doping dependence of $\Delta_0$ in LBCO (triangles) \cite{Valla2006} and LSCO (circles and squares) \cite{Zhou2004,Damascelli2003}.}
\label{Fig7}
\end{center}
\end{figure}

Our study provides the evidence for a $d$-wave gap in the normal ground state of a cuprate
material. Previous studies on underdoped Bi$_2$Sr$_2$CaCu$_2$O$_{8+\delta}$ (BSCCO) were always affected by the superconductivity: The disconnected \lq\lq{}Fermi arcs\rq\rq{} were seen, shrinking in length as $T$ was lowered below pseudogap temperature $T^{\star}$ and collapsing onto (nodal) points below $T_c$ \cite{Norman1998,Kanigel2006,Kanigel2007}.
As a result of this abrupt intervention of superconductivity, it was not clear whether the
pseudogap ground state would have a Fermi arc of finite length or a nodal point or whether it
would be entirely gapped. In LBCO, the absence of superconductivity at $x = 1/8$ has enabled us to show that the normal state gap has isolated nodal points in the ground state. This result points to the pairing origin of the pseudogap. With increasing $T$, a finite-length Fermi arc forms, as suggested in Fig. 5b, in accord with results on BSCCO \cite{Kanigel2006,Kanigel2007}.
\begin{figure}[]
\begin{center}
\includegraphics[width=6.5cm]{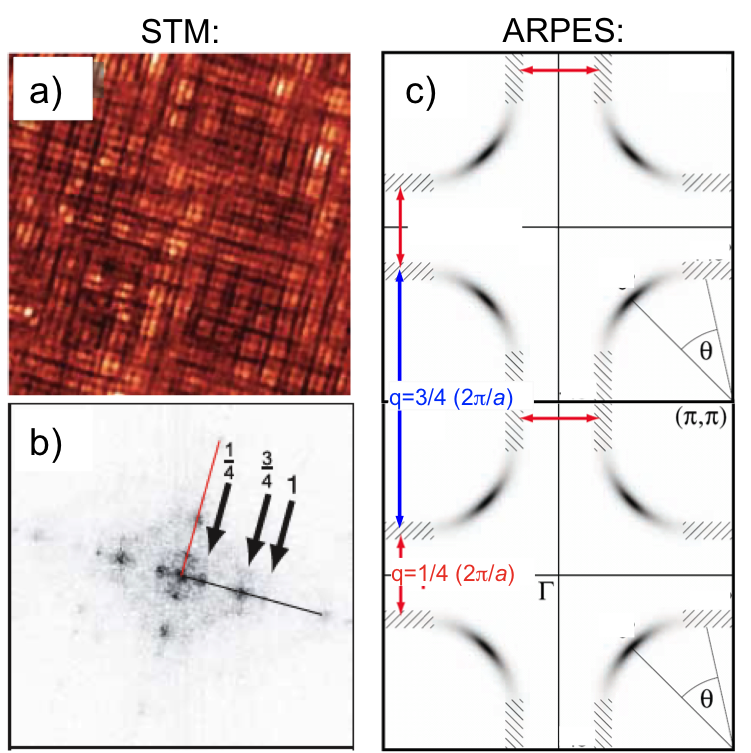}
\caption{STM \cite{Hanaguri2004} and ARPES \cite{Shen2005} on lightly doped Ca$_{2-x}$Na$_x$CuO$_2$Cl$_2$. (a) differential conductance and (b) its Fourier transform in STM reveal a checkerboard superstructure with $q=1/4(2\pi/a_0)$ and $q=3/4(2\pi/a_0)$. (c) ARPES on the same material shows that the Fermi surface can be nested by the same vectors.}
\label{Fig8}
\end{center}
\end{figure}
\begin{figure}[b]
\begin{center}
\includegraphics[width=7cm]{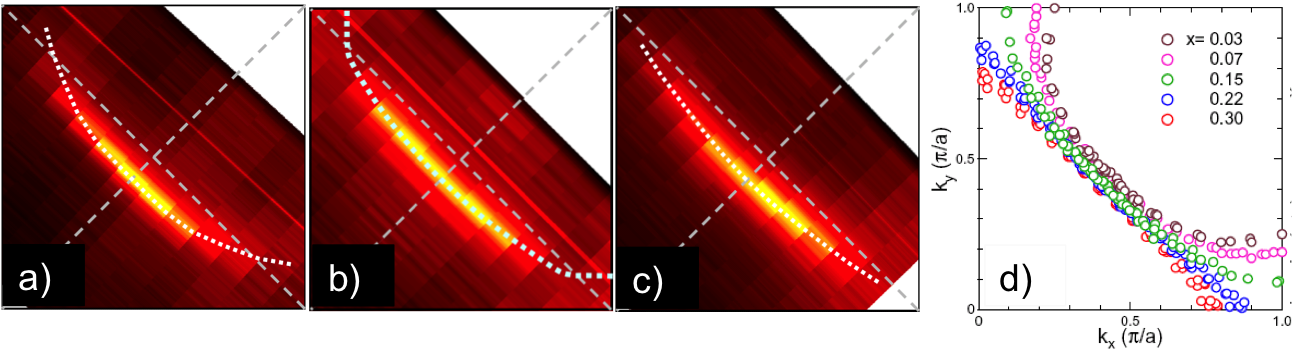}
\caption{Doping dependence of the Fermi surface in 214 cuprates. (a-c) LBCO, at $x=0.095$, $x=1/8$ and $x=0.165$ \cite{Valla2007}. (d) LSCO, for several dopings in the range $0.03<x<0.3$ \cite{Yoshida2006} }
\label{Fig9}
\end{center}
\end{figure}
\begin{figure}[t]
\begin{center}
\includegraphics[width=7cm]{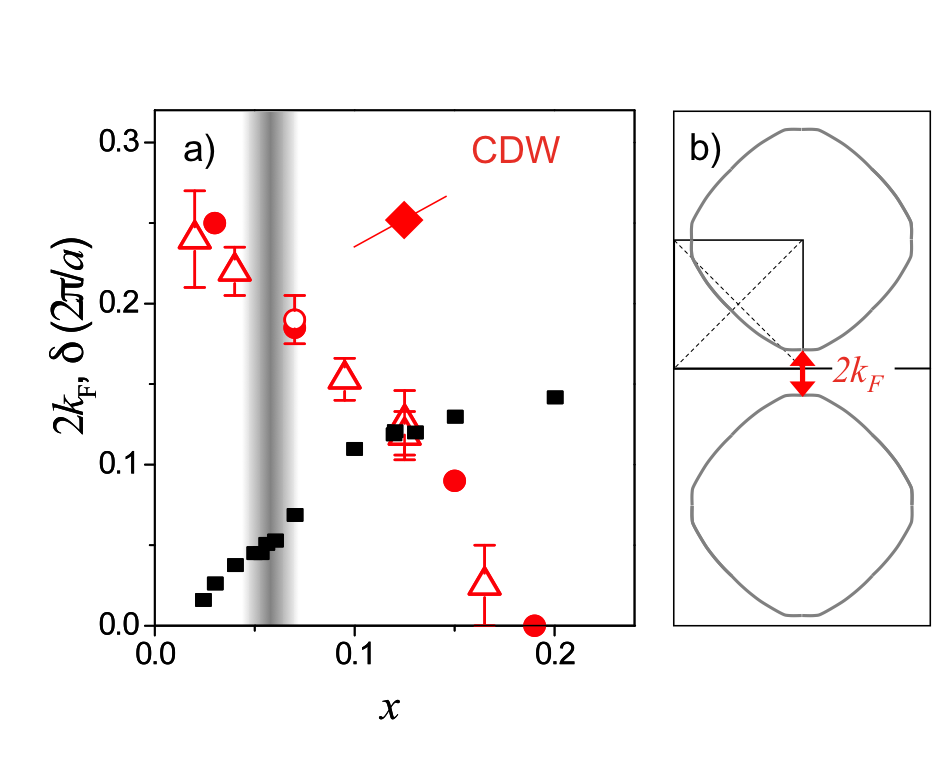}
\caption{A compilation of relevant wave vectors from neutron and x-ray scattering and ARPES on 214 materials. a) Doping dependence of the antinodal $k_F$, as indicated in (b) by the arrow, in LBCO (triangles) \cite{Valla2006} and LSCO (circles) \cite{Yoshida2006}. The wavevector for charge order (diamond) \cite{Abbamonte2005} and the incommensurability $\delta$ from $(\pi,\pi)$ point from neutron-scattering experiments \cite{Fujita2002,Tranquada2004} (squares) are also shown. The gray bar represents the boundary between the \lq\lq{}diagonal\rq\rq{} and \lq\lq{}parallel\rq\rq{} spin superstructures and the onset of superconductivity. (b) A sketch of relevant vectors in the $k$-space.}
\label{Fig10}
\end{center}
\end{figure}

What might be the origin of the observed d-wave gap in LBCO if
superconductivity is absent? Neutron and $x$-ray scattering studies on the same crystal
have identified a static spin order and a charge order \cite{Abbamonte2005,Tranquada2004}. Therefore, it would be tempting to assume that at least a portion of the measured gap is due to the charge/spin order,
in analogy with conventional 2-dimensional (2D) charge-density-wave (CDW) and spin-density wave (SDW) systems.
It has been suggested that in cuprates the charge/spin ordered state forms in a way
where carriers doped into the AF insulator segregate into one-dimensional (1D) charge
rich structures (‘stripes’) separated by the charge poor regions of a parent
antiferromagnet \cite{Emery1999,Tranquada1995,Zaanen1989,Orgad2001}. Some of the predicted spectral signatures of such 1D-like structures would be the states that are relatively narrow in momentum, but broad in energy \cite{Orgad2001}, in a good agreement with the experiments (see Fig. 4 and ref. \cite{Valla1999,Valla2000}, for example). However, questions have often been raised on how to reconcile these unidirectional structures with an apparent 2D Fermi surface and a gap
with $d$-wave symmetry. In the more conventional view, doped carriers are delocalized
in the planes, forming a 2D Fermi surface that grows in proportion with carrier
concentration. The charge/spin ordered state may then be formed in the particle-hole
channel by nesting of Fermi surface segments, producing a divergent electronic
susceptibility and a Peierls-like instability and pushing the system into a lower energy
state with a single particle gap at nested portions of the Fermi surface. 
Therefore, if this is a relevant scenario for the origin of charge/spin order in cuprates, there should be favourable nesting conditions on (portions of) the FS and these conditions would have to change with doping in a way that is consistent with changes observed in (neutron and $x$-ray) scattering experiments.
An example of a cuprate where such a ‘nesting’ scenario is proposed to be at play is Ca$_{2-x}$Na$_x$CuO$_2$Cl$_2$ (CNCOC) \cite{Shen2005}. STM studies have detected checkerboard-like modulations in local DOS on the surface of this material, with $4a\times4a$ periodicity, independent of doping \cite{Hanaguri2004}. Subsequent ARPES studies on the same system have shown a Fermi surface with a nodal arc and truncated anti-nodal segments \cite{Shen2005}. The anti-nodal segments can be
efficiently nested by $q_{CDW}=2k_F=\pi/(2a)$ (and $3\pi/(2a)$) - the same wave vectors observed
in STM for charge superstructure, making the nesting scenario viable. However, as the area enclosed by the Fermi surface in Fig. 8c is much smaller than the nominal doping would require (corresponding to the electron doping), it is suggestive that the nesting might be at play at the surface of CNCOC, but the physics in bulk of this material could be very different. 

If we apply the same nesting scenario to LBCO at $x=1/8$,
we obtain $q_{CDW}\sim4k_F$ $(=\pi/2a)$, for charge order, instead of $2k_F$ nesting, suggested to be at
play in CNCOC. Moreover, from the doping dependence of the Fermi surface in both LBCO and LSCO (Fig. 9), the nesting of anti-nodal segments would produce a wavevector
that shortens with doping, opposite of that observed in neutron scattering studies
in terms of magnetic incommensurability. This is illustrated in Fig. 10, where we compile
the doping dependences of the anti-nodal $2k_F$ for LBCO \cite{Valla2006} and LSCO \cite{Yoshida2006} samples and wave-vectors measured in scattering experiments \cite{Fujita2002,Tranquada2004,Abbamonte2005}.
There is another, more fundamental problem with the \lq{}nesting\rq{} scenario: any
order originating from nesting (particle-hole channel) would open a gap only on nested
segments of the Fermi surface, preserving the non-nested regions. The fact that only
four gapless points (nodes) remain in the ground state essentially rules out nesting as an
origin of pseudogap. In addition, a gap caused by conventional charge/spin order would
be pinned to the Fermi level only in special cases \cite{Schafer1999}. The observation that it is always
pinned to the Fermi level (independent of $k$-point, as measured in ARPES and of doping
level, as seen in STM on different materials) and that it has $d$-wave symmetry
undoubtedly points to its pairing origin - interaction in the particle-particle singlet
channel. Note that, in contrast to the low-energy pairing gap, STM at higher
energies shows a DOS suppressed in a highly asymmetric manner, indicating that some
of the \lq{}nesting\rq{}-related phenomena might be at play at these higher energies (Fig. 6b).

The surprising anti-correlation of the low-energy pairing gap and $T_c$ over some
region of the phase diagram suggests that in the state with strongly bound Cooper pairs,
the phase coherence is strongly suppressed by quantum phase fluctuations. Cooper pairs
are then susceptible to spatial ordering and may form various unidirectional \cite{Tranquada1995,Emery1999} or 2D \cite{Tesanovic2004,Hanaguri2004} superstructures. Quantum phase fluctuations are particularly prominent in cases where such superstructures are anomalously stable. Some of these effects have been observed in transport properties \cite{Li2007a,Tranquada2008}, but the connection to the specific models has not been firmly established. For some of the theoretically proposed structures, the quantum phase fluctuations are strongest at the doping of $1/8$, in general agreement with our
results: $1/8$ represents the most prominent \lq{}magic fraction\rq{} for a checkerboard-like
\lq{}CDW of Cooper pairs\rq{} \cite{Tesanovic2004}, and it locks the \lq{}stripes\rq{} to the lattice in a unidirectional alternative. The presence of nodes in the ground state of the pseudogap represents a new
decisive test for validity of models proposed to describe such structures.
 
The more recent ARPES studies \cite{Tanaka2006,He2008} have suggested that the apparent deviations from the perfect (cos$k_x-$cos$k_y$)/2 dependence of the single-particle gap in some underdoped cuprates point to the co-existence of two gaps: the pairing one, that resides near the node and scales with $T_c$ and the second one, that originates from the competing charge/spin orders, resides near the anti-nodes and scales with $T^\star$. We note that in the case of LBCO, this picture does not seem to make much sense, since $T_c\rightarrow0$ would imply that the near-nodal, \lq{}pairing\rq{} gap also goes to zero, opposite of the experimental observation \cite{Valla2006,He2008}.


\section{Diagonal Stripes}
\label{Dia}
\begin{figure}[]
\begin{center}
\includegraphics[width=8.5cm]{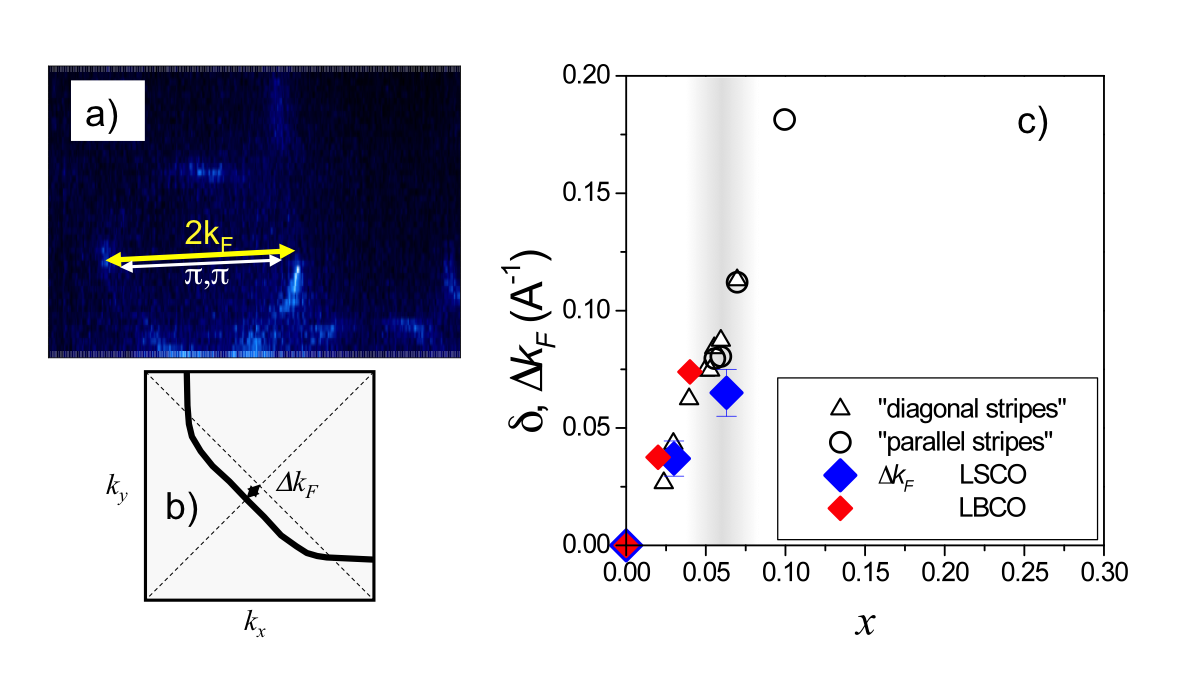}
\caption{(a) LBCO Fermi surface at $x=0.04$. The nodal $k_F$ is measured from $(\pi/2,\pi/2)$ point as indicated in (b) and its doping dependence in the \lq\lq{}spin glass\rq\rq{} regime is shown in (c) (red diamonds). Data for LSCO from ref. \cite{Zhou2004,Yoshida2006} are represented by blue diamonds. Also shown is the incommensurability $\delta$ measured in neutron scattering close to the onset of superconductivity (gray area) where the spin structure rotates from diagonal ($x<0.06$, open triangles) to parallel ($x>0.06$, open circles) direction relative to the Cu-O bond.}
\label{Fig11}
\end{center}
\end{figure}

There is another important evidence for a close relationship between superconductivity and static charge/spin superstructures at $x=0.055$ where superconductivity just appears.
There, a peculiar transition has been also observed in neutron
scattering experiments on LSCO samples \cite{Fujita2002,Wakimoto2000,Wakimoto2001}: a static incommensurate scattering near AF wave vector with incommensurability $\delta\propto x$ rotates by $45^\circ$, from being diagonal to Cu-O bond ($x\leq0.055$) to being parallel to it
($x\geq0.055$). More recently, similar features were also observed in LBCO, with the similar doping dependence, indicating the common nature of this transition in the 214 family of cuprates \cite{Dunsiger2008}. The transition from diagonal (static) to parallel (dynamic) \lq{}stripes\rq{} coincides with the transition seen in transport \cite{Ando2001} where "insulator" turns into superconductor for $x>0.055$. Although the in-plane transport becomes "metallic" at high temperatures even at 1\%
doping, at low temperature there is an upturn in $\rho_{ab}$, indicating some
localization. This upturn is only present for $x<0.055$, where the diagonal spin incommensurability exists.
The moment diagonal points in the neutron scattering experiments disappear (rotate), the upturn in resistivity vanishes, and superconductivity appears. 

The early ARPES experiments on LSCO in this low doping regime \cite{Zhou2004,Yoshida2003} have shown that the first states appearing at the Fermi level are those near the nodal line, whereas the rest of the Fermi surface is affected by a large gap of similar $d$-wave symmetry as the superconducting gap. 
A closer look at ARPES results from this region of the phase diagram uncovers that the diagonal incommensurability $\delta$ seen in neutron scattering is closely related to the increase in $k_F$ of nodal states \cite{Zhou2004,Yoshida2003}. 
\begin{figure}[]
\begin{center}
\includegraphics[width=7.5cm]{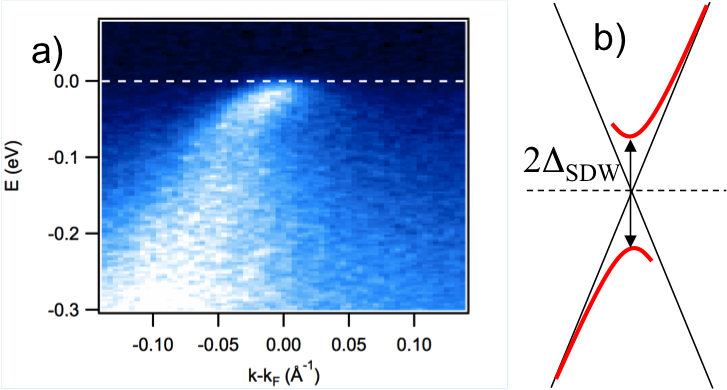}
\caption{a) ARPES spectrum from the nodal line of the $x=0.04$ LBCO sample. A small gap ($\Delta\approx3$ meV) indicates that the \lq\lq{}diagonal\rq\rq{} stripes act as the conventional spin-density-waves (SDW). b) SDW gap}
\label{Fig12}
\end{center}
\end{figure}
In Fig. 11, we show the Fermi surface of an LBCO sample at $x=0.04$. The nodal span of the Fermi surface, $2k_F$, that can be precisely extracted from these measurements, shows the same doping dependence as the diagonal incommensurability from neutron scattering. Our results from $x=0.02$ and $x=0.04$ samples show that $\delta=k_F-\pi/2$ in this low-doping regime, as can be seen from Fig. 11c. At higher doping levels, the Fermi surface grows more in the anti-nodal region, while the nodal $k_F$ saturates near $x\approx0.07$ (Fig. 9(a-d)), similar to the evolution seen in LSCO \cite{Zhou2004}. The relationship $\delta=k_F-\pi/2$ in the \lq{}spin-glass\rq{} phase suggests that the diagonal incommensurate scattering may be understood in terms of conventional
spin-density wave (SDW) picture, originating from 2$k_F$ nesting of nodal
states. Such SDW would open the gap and localize the nodal states at low
temperatures, preventing the superconductivity from occurring, in agreement with
transport \cite{Ando2001,Ando2004} and optical \cite{Homes2006,Dumm2003} studies. 
Indeed, we have detected a small gap ($\Delta\approx3$ meV) at the nodal point of the $x=0.04$ sample, as can be seen in Fig. 12, leaving the whole Fermi surface gapped (Fig. 13). Similar gap has been also observed in LSCO samples at similar doping levels \cite{Shen2004} This reaffirms our proposal that the nodal states are affected by the \lq\lq{}diagonal\rq\rq{} stripes in a similar way as they would be in the case of the conventional, Fermi-surface nesting induced SDW. The observed nodal SDW (or \lq{}diagonal stripe\rq{}) gap has dramatic consequences on the transport properties in this doping regime of the phase diagram.
While it is clear that the near-nodal states are responsible for metallic normal state
transport at these low doping levels, their role in superconductivity is not
adequately appreciated. We think that they actually play a crucial role in superconductivity itself: when the diagonal SDW vanishes, the nodal states are released and
superconductivity follows immediately. 
The role of near nodal states in superconductivity might be only a secondary one: the one of phase coherence propagators, where due to the small superconducting gap $\Delta(k)$, these states have large coherence length $\xi(k)=v_F(k)/|\Delta(k)|$, and are able to stabilize the global
superconducting phase \cite{Joglekar2004} in the presence of inhomogeneities observed in
STM in underdoped Bi$_2$Sr$_2$CaCu$_2$O$_{8+\delta}$ \cite{Hoffman2002,Pan2001,Lang2002}. However, it might as well be that the superconductivity is a relatively weak order parameter
residing only in the near-nodal region, disrupted by something much larger, with similar symmetry that dominates the rest of the Fermi surface in the underdoped regime \cite{Varma1999,Chakravarty2001,Bernevig2003}.
\begin{figure}[]
\begin{center}
\includegraphics[width=6cm]{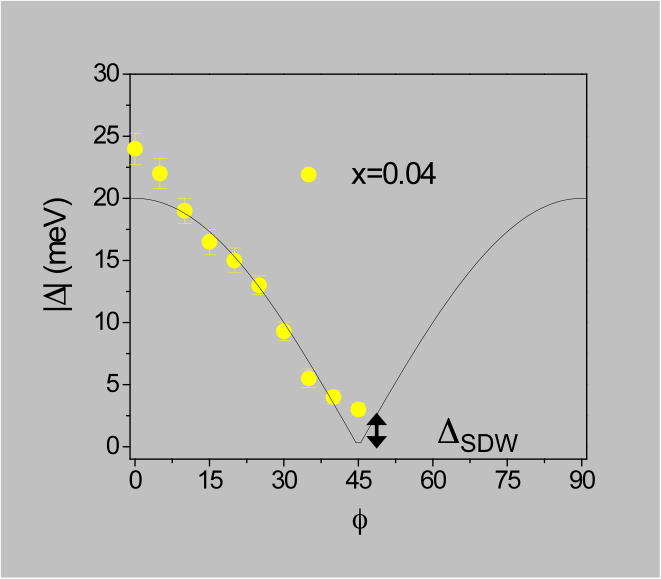}
\caption{Magnitude of single-particle gap in LBCO at $x=0.04$ as a function of an angle around the Fermi surface at $T=12$ K. The whole Fermi surface is gapped at this doping level. }
\label{Fig13}
\end{center}
\end{figure}

\section{Conclusions}
\label{Conclusions}
In conclusion, I have pointed out that experimental results on cuprate superconductors indicate a very strong connection between spin susceptibility measured in neutron scattering and single-particle properties measured in ARPES and superconductivity itself. The interplay of spin fluctuations and superconductivity is particularly evident at the onset ($x\approx0.05$) of superconductivity, where the static “\rq\rq{}diagonal\rq\rq{} incommensurate spin density wave gives away to superconductivity and at $x=1/8$, where the freezing of spin fluctuations and the appearance of static \lq\lq{}parallel\rq\rq{} stripes suppresses superconducting transition temperature almost to zero.
We have found that in the ground state of LBCO at $x=1/8$, a system with static spin and charge  orders \cite{Abbamonte2005,Tranquada2004} and no superconductivity, the $k$-dependence of the single-particle gap looks the same as the superconducting gap in superconducting cuprates: it has magnitude consistent with $d$-wave symmetry and vanishes at four nodal points on the Fermi surface. Furthermore, the gap, measured at low temperature, has a doping dependence with a maximum at $x\approx1/8$, precisely where the charge/spin order is established between two adjacent superconducting domes. These findings reveal the pairing origin of the \lq\lq{}pseudogap\rq\rq{} and imply that the most strongly bound Cooper pairs at $x\approx1/8$ are most susceptible to phase disorder and spatial ordering \cite{Emery1995,Tranquada1995,Tesanovic2004}.
In the low-doping regime ($x<0.05$), we have uncovered the tight relationship between the \lq{}diagonal stripes\rq{} and the single-particle spectral features in the near-nodal region of the Fermi surface in 214 cuprates. The diagonal super-structures or \lq{}diagonal stripes\rq{}, seen in neutron scattering when the first carriers are doped into a parent compound, originate from the Fermi surface nesting of the nodal segments in a conventional SDW manner. Further doping above $x\approx0.055$ destroys these superstructures and releases the nodal states which then can play a role in establishing the superconducting phase coherence.
\section{acknowledgments}
\label{ack}
I would like to acknowledge useful discussions with John Tranquada, Peter Johnson, Chris Homes, Sa\v sa
Dordevi\'c, Myron Strongin, Alexei Tsvelik, Steve Kivelson, Doug Scalapino, Alexander Kordyuk, Genda Gu, Shuichi Wakimoto, Seamus Davis, Zlatko Te\v sanovi\'c, and Atsushi Fujimori. The program was supported by the US DOE under contract number DE-AC02-98CH10886.



\bibliographystyle{elsarticle-num}

\end{document}